\documentclass[a4paper,11pt]{article}
\usepackage{pos}
\usepackage{float}

\title{Early Time Dynamics and the Bulk}
\author{Bin Wu\footnote{Speaker's slides: \url{https://indico.cern.ch/event/751767/contributions/3840642/attachments/2048915/3433716/2020.pdf}}}

\affiliation{Theoretical Physics Department, CERN\\ CH-1211 Geneva 23, Switzerland}

\emailAdd{b.wu@cern.ch}

\abstract{Deciphering the origin of collective phenomena in small colliding systems is one of contemporary focuses in heavy-ion physics. It entails penetrating the barrier between two previously separated research topics: thermalization/hydrodynamization and phenomenological studies of collectivity. I first review some recent progress in understanding thermalization/hydrodynamization in large colliding systems, centralized on bottom-up thermalization. Then, using a simple kinetic theory I demonstrate how the investigation of hydrodynamization is intertwined with the study of flow in small colliding systems. Connections of these studies to "hard probes" are also commented where possible.}

\FullConference{%
  HardProbes2020\\
  1-6 June 2020\\
  Austin, Texas}

\begin{document}
\maketitle

\section{Introduction}

This proceeding focuses on physical pictures behind early time dynamics of QCD bulk matter produced in high-energy hadronic/nuclear collisions. To be exact, "early time" here means any time before hydrodynamic modes start to dominate in bulk matter, which may never happen in small colliding systems. Accordingly, this proceeding consists of two parts: a review on the formation of a QGP fluid droplet in large colliding systems and a report on the connection between flow and hydrodynamization in small colliding systems.

This proceeding is confined to weakly-coupled quantum field theories, especially perturbative QCD. All the discussions are directly based on, or carrying some features of, the parton picture as first discussed by Bjorken~\cite{Bjorken:1976mk}: QCD bulk matter first emerges as "wee" partons radiated in binary collisions of valence quarks of the two colliding objects. The valence quarks mostly pass through each other and carry away the Baryon numbers into forward rapidities. Due to parton saturation~(see \cite{Kovchegov:2012mbw} for a diactic introduction), these wee partons typically carry an energy of order $Q_s $ (the saturation momentum), which becomes a perturbative scale at RHIC and LHC energies. 
Because soft gluon radiation is independent of valence quarks' energies, bulk matter features invariance with respect to longitudinal boosts, filling a central plateau in rapidity. Based on this picture, longitudinal boost-invariance is posited in all the following discussions.

This proceeding is organized as follows. Section \ref{sec:AA} is devoted to a brief review on progress in the study of thermalization/hydrodynamization in large colliding systems, focusing on bottom-up thermalization~\cite{Baier:2000sb}. A review on flow in small colliding systems as an intricate interplay between hydrodynamic and non-hydrodynamic modes of QCD  bulk matter is given in Sec. \ref{sec:small}. Some comments pertinent to the pivotal topics on hard probes of this conference are also given wherever possible.

\section{Thermalization in Large Colliding Systems}
\label{sec:AA}
In a central AA collision, the thermalization time is expected to be much shorter than its transverse size. Since the transverse expansion is negligible compared to the longitudinal expansion, one essentially winds up with a 1+1D (one spatial and one time dimension) system. In this case, wee partons turns out to establish thermal equilibrium in a "bottom-up"  fashion by quenching all the wee partons with energy of order $Q_s$ and, hence, heating up a thermal QGP bath to its maximal temperature~\cite{Baier:2000sb}.

\subsection{Parton energy loss in longitudinally boost-invariant plasma}
The details of thermalization are governed by the physics on how partons are quenched in a QGP via medium-induced energy loss. The longitudinally boost-invariant plasma only rarefies notably within a period of time $\Delta\tau\gtrsim \tau$ with $\tau$ the proper time. At this time radiation with formation time $\tau_f\ll \tau$ witnesses no significant change in the properties of the plasma and, accordingly, the radiated gluon spectrum reduces to the static case with the jet quenching parameter replaced by a time dependent one $\hat{q}(\tau)$:
    \begin{align}
        \omega\frac{dI}{d\omega} \sim \alpha_s N_c \sqrt{\frac{\hat{q}(\tau)\tau^2}{\omega}}\equiv \alpha_s N_c\frac{\tau}{\tau_{f}}
        \qquad\text{for $\tau_f \ll \tau$}.
    \end{align}
This is the main result for radiative energy loss that helps us to understand bottom-up thermalization. It can be easily checked by taking this limit of the full result in \cite{Baier:1998yf} (see, also, \cite{Arnold:2008iy}). And another demonstration of the validity of this time scale argument is the logarithmically enhanced radiative correction to $p_T$-broadening and $\hat{q}$. Detailed calculations show that it is the same as a static medium with $\hat{q} = \hat{q}(\tau)$ at leading logarithmic accuracy~\cite{Iancu:2018trm}
. 

\subsection{Bottom-up thermalization}

The coherence time of wee partons is evidently given by $1/Q_s$. At $\tau\sim 1/Q_s$, they are expected to lose coherence and start to behave like particles roughly on mass-shell. As a quantitative understanding of such a physical argument, one can carry out an analytical calculation of wee partons' two point correlation function and show that at tree level~\cite{Wu:2017rry}
\begin{align}\label{eq:fcl}
G_{22}^{a\mu,b\nu}(X, p) \to  2\pi\delta(p^2) \delta^{ab}
  \sum\limits_{\lambda=\pm}\epsilon_\lambda^\mu(p)
  \epsilon_{\lambda}^{*\nu}(p) f^{cl}(X,p)\qquad\text{as $\tau\to\infty$},
\end{align}
where $f^{cl}(X,p)\propto \delta(\tau p_z)$ is a quasi-classical distribution and the $\delta$ funciton imposes the on-shell condition of quasiparticles. This result also confirms that the  typical longitudinal momentum scales like $1/\tau$ under free-streaming (when interactions among wee partons can be neglected). The quasi-particle distribution at $\tau\sim 1/Q_s$ is saturated \cite{JalilianMarian:1996xn, Kovchegov:1998bi, Mueller:1999wm}:
\begin{align}
    f\sim \frac{1}{\alpha_s}\qquad\text{for $p\lesssim Q_s$}.
\end{align}
Due to longitudinal boost-invariance, one needs only to study the system at one longitudinal location. In all the following discussions, this location is chosen to be at $z=0$, where the collision occurs.

Afterwards, inelastic scattering among wee gluons populates the softer sector of the gluon distribution. In order to avoid the confusion of wee and softer gluons, they are respectively called hard and soft gluons below. The system establishes thermal equilibrium via interactions among hard and soft gluons under longitudinal expansion. It undergoes three distinctive stages in the limit $\alpha_s\ll 1$~\cite{Baier:2000sb}, which has been confirmed by numerical simulations using kinetic theory~\cite{Kurkela:2015qoa}. Below, I reiterate some main points of this thermalization scenario in terms of $\hat{q}$, one of the most important parameters for studying jet quenching:

\noindent {{\bf Stage I} ( $\alpha_s^{-\frac{3}{2}}> \tau Q_s > 1$): expansion prevails interaction.}\\
    At this stage, soft gluons play no dominant roles in any physical effects. The jet quench parameter scales like
    \begin{align}
    \hat{q}\sim \alpha_s^2 N_h f_h \sim Q_s^3/(\tau Q_s)^{5/3},
    \end{align}
    where $N_h\sim Q_s^2/(\alpha_s \tau)$ is the number density of hard gluons and $f_h\sim 1/(\alpha_s \tau p_z)\gg 1$ is their phase-space distribution. Since hard gluons with $p_z \sim Q_s$ escape rapidly from the transverse plane at $z=0$, the non-vanishing longitudinal momentum of the remainder is predominantly given by multiple scattering, that is, $p_z^2 =\hat{q}\tau\sim Q_s^2/(\tau Q_s)^{2/3}$. As a result, the pressure anisotropy
    \begin{align}
        P_L/P_T\sim p_z^2/p_T^2 \sim 1/(\tau Q_s)^{2/3}.
    \end{align}
    This scaling behavior has been confirmed numerically using classical statistical field simulations and identified as a universal attractor~\cite{Berges:2013fga}. 
    
\noindent
Generically, one can expect such a decrease of $P_L/P_T$ in kinetic theory as long as the longitudinal expansion dominates (see also \cite{Epelbaum:2015vxa} for $\phi^4$ theory and \cite{Kurkela:2019set} for kinetic theory in isotropization time approximation (ISA)). This distinguishes weakly-coupled systems from strongly-coupled systems, which apparently lack such a universal behavior~\cite{Beuf:2009cx, Wu:2011yd}. The quantitative study in ~\cite{Kurkela:2019set} shows that kinetic field theory (in ISA), Israel-Stewart (IS) hydrodynamics and AdS/CFT differ most distinctively in their early-time attractors.
    
\noindent
{{\bf Stage II} ($ \alpha_s^{-\frac{5}{2}}> \tau Q_s > \alpha_s^{-\frac{3}{2}}$): interaction countervails expansion.}\\
At this stage, soft gluons starts to contribute dominantly to Debye screening although the number of soft gluons is still less than that of hard ones. Since $f_h$ becomes smaller than 1, one has
    \begin{align}
        \hat{q}\sim \alpha_s^2 N_h \sim\alpha_s Q_s^2/\tau\Rightarrow p_z^2 = \hat{q}\tau \sim \alpha_sQ_s^2,
    \end{align}
    and, accordingly,
    \begin{align}
        P_L/P_T\sim p_z^2/p_T^2 \sim \alpha_s.
    \end{align}
    At $\tau \sim \alpha_s^{-5/2}/Q_s$, the number density of soft gluons becomes parametrically equal to $N_h$. From the soft gluon energy density $\epsilon_s\sim (\hat{q}\tau)^{\frac{1}{2}} N_h$, one can see that their thermalization time commences to be comparable to $\tau$:
    \begin{align}
        \frac{\tau_{th}}{\tau} \sim \frac{1}{\alpha_s^2 \epsilon_s^{\frac{1}{4}}\tau}\sim 1.
    \end{align}
    Now, soft gluons are poised to form a thermal bath.

\noindent {{\bf Stage III} ($\alpha_s^{-\frac{13}{5}}> \tau Q_s > \alpha_s^{-\frac{5}{2}}$): quenching of "hard" gluons in a thermal bath of soft gluons.}\\
    At his stage, soft gluons outnumber the hard ones and form a thermal bath. So thermalization literally proceeds as jet quenching.
    In this case, the typical radiative energy loss of a hard gluon is given by $\Delta E\sim \alpha_s^2 \hat{q} \tau^2$ \cite{Baier:2001yt} and the jet quenching parameter $\hat{q}\sim \alpha_s^2 \epsilon_s^{\frac{3}{4}}$. Employing these formulas, one can easily estimate the energy density of the thermal bath
    \begin{align}
        \epsilon_s \sim N_h \Delta E \Longleftrightarrow T \sim \epsilon^{\frac{1}{4}}\sim \alpha_s^3Q_s^2 \tau\text{ and }\hat{q}\sim \alpha_s^{11} Q_s^3(Q_s\tau)^3.
    \end{align}
    Accordingly,
    \begin{align}
        P_L/P_T\sim \frac{\epsilon_s}{N_h Q_s}\sim \alpha_{s}^{13} (\tau Q_s)^{5}.
    \end{align}
    At $\tau\sim \alpha_s^{-\frac{13}{5}}/Q_s$, $P_L/P_T\sim 1$ and $\epsilon_s\sim N_h Q_s$, signaling the establishment of thermal equilibrium.

\begin{figure}[H]
    \centering
    \includegraphics[height=0.2\textheight]{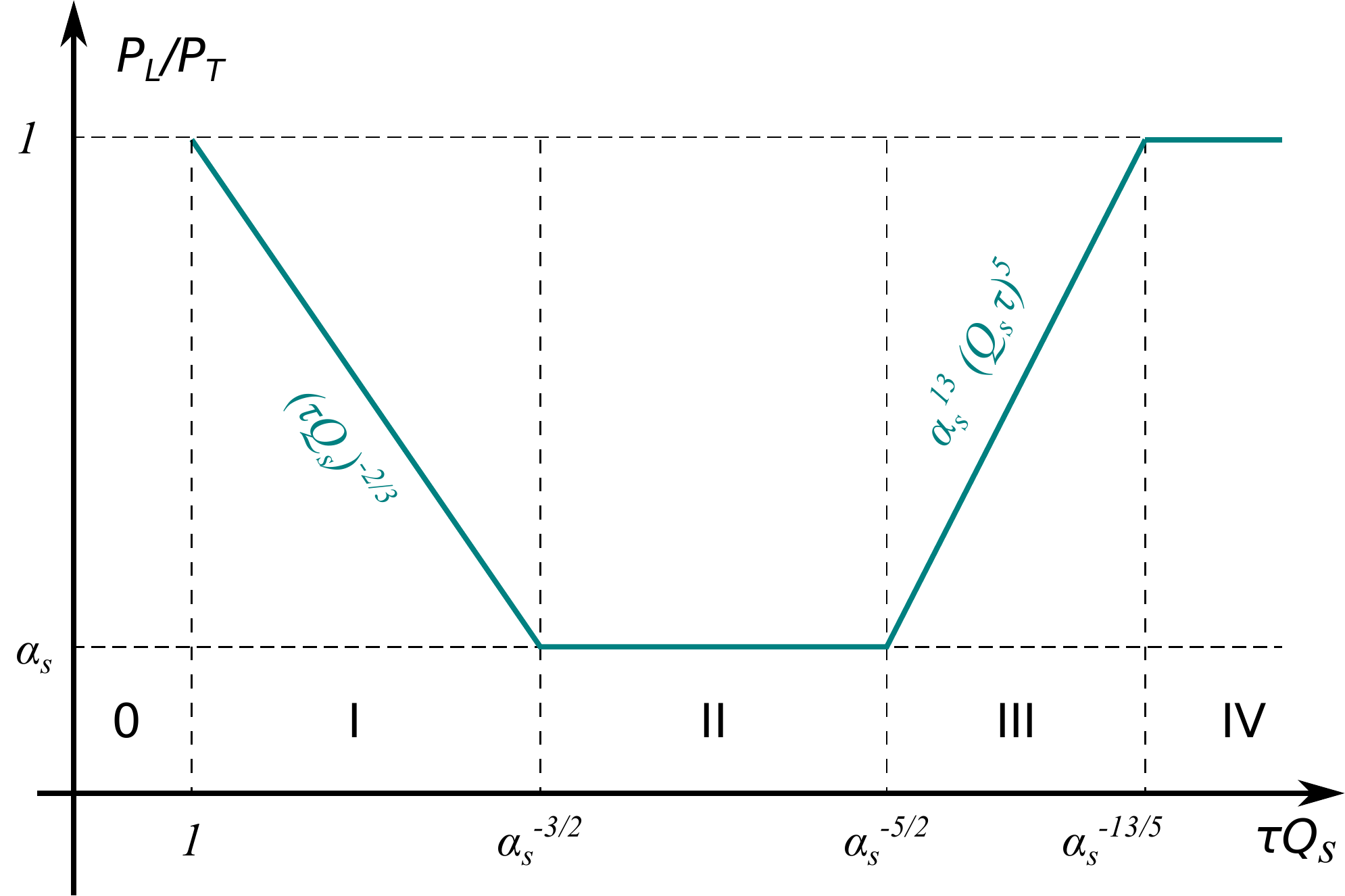}\hspace{0.01\textwidth}\includegraphics[height=0.2\textheight]{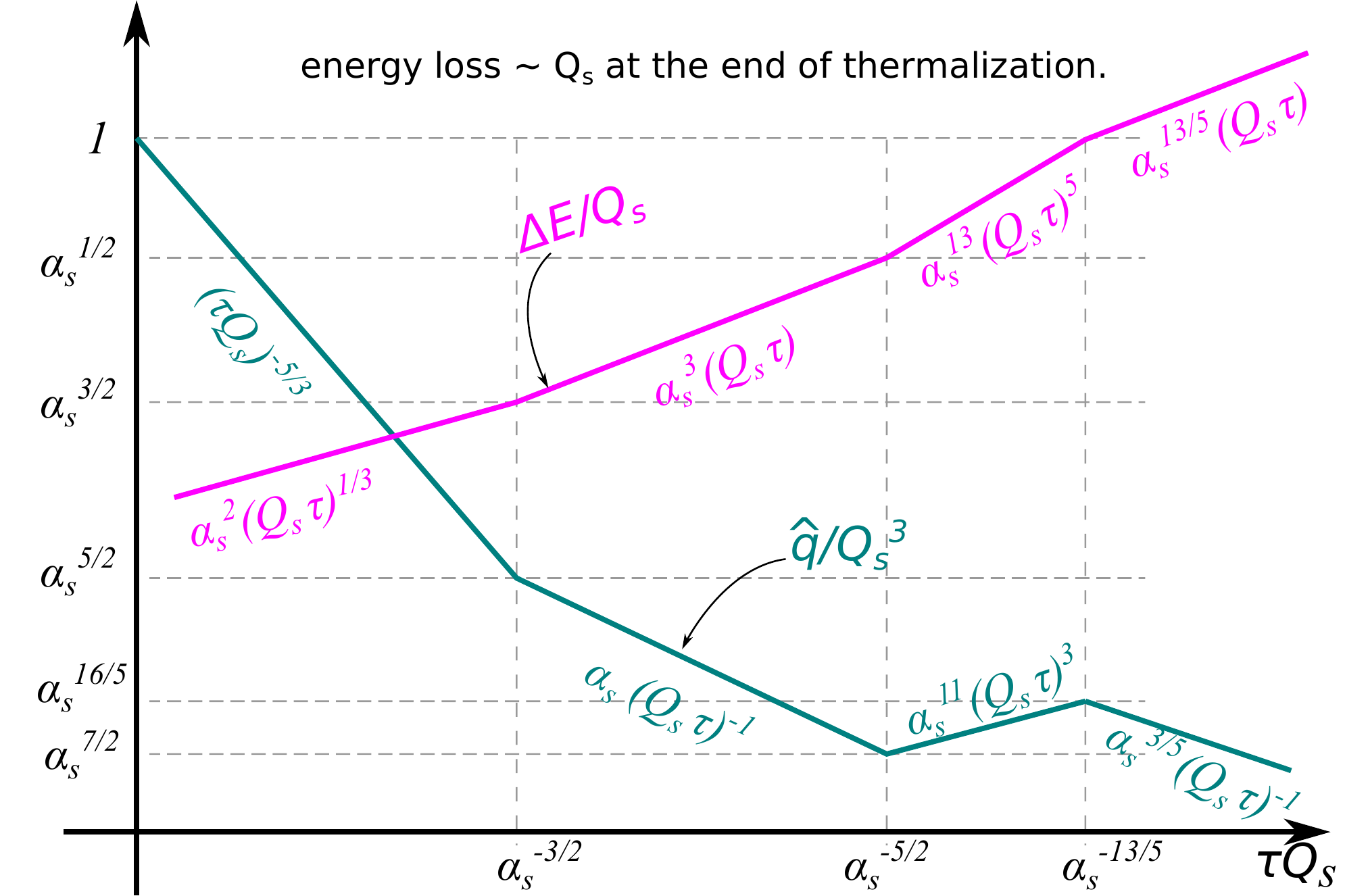}
    \caption{Pressure anisotropy $P_L/P_T$ (left) and $\hat{q}$ and radiative energy loss (right).}
    \label{fig:bottomup}
\end{figure}

Fig. \ref{fig:bottomup} shows the pressure anisotropy and the information relevant for jet quenching based on the above discussion. One would expect that hard QCD jets also lose an energy of order $Q_s$ during the thermalization process.

\subsection{Attempts to go beyond bottom-up thermalization}

Parametrically, one expects that the tree level diagrams connected to binary scattering of valence quarks give dominant contributions before the onset of Stage II at $\tau\lesssim \alpha_s^{-\frac{3}{2}}/Q_s$. This group of diagrams can be summed over by solving classical Yang-Mills equation. Starting from Stage I for $\tau \gtrsim 1/Q_s$, another group of  diagrams corresponding to two point irreducible (2PI) diagrams in the effective action are needed to be summed over in order to describe all the three stages of the thermalization process. Kinetic theory can be viewed as an approximation to the sum of this second group of diagrams. Parametrically, the diagrams common to these two groups give the dominant contributions during Stage I. Therefore, in practice one may switch between classical field simulations to kinetic theory at some time during this stage based on the assumption in which quasiparticle approximation is justified~\cite{Mueller:2002gd}. The interested reader is referred to a recent implementation of such a scheme~\cite{Kurkela:2018wud}.

Back to quantum field theory, there are some interesting theoretical issues to scrutinize:

\noindent{1. Partial quantum effects in statistical classical field theory.} \\
If one includes quantum fluctuations in vacuum at the initial time in classical field simulations, one can actually go beyond the classical thermal field point of the from $f=T/p$ for bosons~\cite{Mueller:2002gd, Jeon:2004dh, Epelbaum:2014mfa}. Including such partial quantum effects leads to the observation of fast pressure isotropization~\cite{Gelis:2013rba}. However, there is a nagging issue about how to deal with ultraviolet (UV) divergences in this approach, which is a non-renormalizable field theory~\cite{Epelbaum:2014yja}. A proposal to deal with such UV divergences can be found for $\phi^4$ theory in \cite{Epelbaum:2014yja}.

\noindent{2. Kinetic theory vs Feynman diagrams in perturbative QCD.}\\
So far there is no unique tool for summing both groups of diagrams. This means that the transition between classical field theory to kinetic theory is mostly based on parametric argument and physical intuition. One available approach to test such a transition is to perform perturbative calculations and investigate such a transition order by order in the coupling~\cite{Wu:2017rry}.  One of the approximation is needed to derive the Boltzmann equation from the second group of diagrams is to replace each appearance of two point functions by a product of the delta function imposing the on-shell condition of particles and a quasi-classical distribution. This, however, is only rigorously true in an infinite period of time, as shown in Eq. (\ref{eq:fcl}). A detail calculation in \cite{Kovchegov:2017way} actually shows that such an approximation is not valid at the lowest order in $\phi^4$ theory. Such a negative result opens up a vista about the richness of quantum field theory, which could not be captured by bottom-up scenario based on the quasi-particle picture. 

\section{Flow in Small Colliding Systems}
\label{sec:small}
In this section, using a simple kinetic theory I demonstrate that both concepts and technical tools for studying thermalization/hydrodynamization are essential to pinning down the origin of flow in small colliding systems. 

\subsection{Hydrodynamic and non-hydrodynamic modes in bulk matter}

QCD bulk matter contains both hydrodynamic and non-hydrodynamic modes. And thermalization/hydrodynamization can be viewed as a process in which hydrodynamic modes eventually prevail after non-hydrodynamic modes phase out. It involves an intricate interplay between hydrodynamic and non-hydrodynamic modes, which, as we shall see, both contribute to collective flow. The study of collectivity in small colliding systems provides us golden opportunities to study such an interplay~\cite{Kurkela:2018qeb}, hence foreshadowing experimental tests of the mechanism underlying thermalization/hydrodynamization.

All know interacting quantum field theories contain hydrodynamics but they differ in non-hydrodynamic modes, meaning they go beyond hydrodynamics in different ways. Non-hydrodynamic modes relevant for the parton picture under consideration are particle-like excitations (quasi-particles), including, e.g., partonic constituents of high-$p_T$ jets, and 
wee partons (hard gluons) studied in bottom-up thermalization. Such a categorization has another advantage compared to the conventional one in which produced particles are divided into bulk matter and hard probes by introducing a somewhat arbitrary momentum scale. It can hence serve as a building block of a unified framework for hard processes in high-energy hadronic/nuclear collisions.

\begin{figure}[H]
    \centering
    \includegraphics[width=0.6\textwidth]{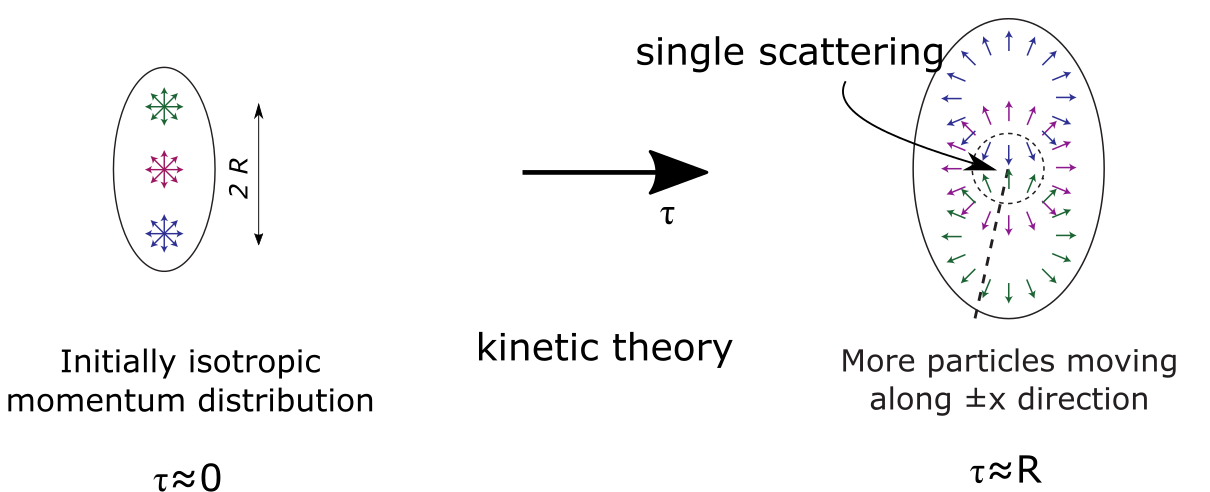}
    \caption{Elliptic flow from single scattering of particle-like excitations~\cite{Kurkela:2018ygx}.}
    \label{fig:onehit}
\end{figure}

Non-hydrodynamic modes (quasi-particles) contribute to elliptic flow in a way very different from hydrodynamic modes~\cite{Kurkela:2018ygx}. As illustrated in Fig. \ref{fig:onehit}, imagine that a small system in question initially possesses a spatial anisotropy while its constituents are isotropically distributed in transverse momentum. The final-state transverse momentum anisotropy is then mostly generated by scatterings in the center at time $\tau\sim R$. Around this time, two particles, initially separated in space by a distance $\sim 2R$ (along the y-axis), encounter and scatter near the center. There is always a chance for these particles, initially moving along the y-axis, to pick up a non-vanishing x-component of their transverse momenta as a result of collision. This is how elliptic flow is generated by single scattering among quasi-particles, which is also responsible for the pressure isotropization.

\subsection{Hydrodynamization and phenomenological studies of flow}

Unlike central AA collisions, one can no longer neglect transverse expansion in small colliding systems and has to deal with a 1+3D problem. Below, I use a simple kinetic theory~\cite{Kurkela:2018qeb} (similar to that in \cite{Baym:1984np}) to demonstrate how flow data can be utilized to discern physics underlying the interplay between hydrodynamic and non-hydrodynamic modes of bulk matter. The exploration using  the full QCD kernel in parallel with bottom-up thermalization is yet to be carried out.

When one uses hydrodynamic models, one has to assure that their predictions are dominated by hydrodynamic modes. All hydrodynamic models include something non-hydrodynamic to insure consistency, which, however, does not correspond to correct non-hydrodynamic modes in underlying quantum field theories. One way to qualify the dominance of hydrodynamic modes is to use ''fluid quality''~\cite{Kurkela:2019kip}
\begin{align}
    Q(t,r) =  \sqrt{  \frac{ \left( T- T_{\rm hyd} \right)^{\mu\nu}   \left( T - T_{\rm hyd} \right)_{\mu\nu} }{ \left(T_{\rm id} \right)^{\mu\nu}    \left(T_{\rm id} \right)_{\mu\nu}    }  },
\end{align}
and define the hydrodynamic dominance by picking some small value, say,
\begin{align}
Q<0.1,
\end{align}
where $T_{\text{hyd}}$ and $T$ are the energy-momentum tensors respectively calculated using the constitutive fluid-dynamic relation and calculated in the theory under investigation. By implementing the  above criteria with $T_{\text{hyd}}$ up to second order in fluid dynamic gradients, one can find that in the kinetic theory non-hydrodynamic modes (particle-like excitations) dominate for $\hat{\gamma}\lesssim 2$ while hydrodynamic modes dominate for $\hat{\gamma}\gtrsim 4$~\cite{Kurkela:2019kip}. Here, opacity $\hat{\gamma}$, the unique parameter of this kinetic theory, is given by the ratio of the system size and the mean free path.

\begin{figure}[H]
    \centering
    \includegraphics[height=0.2\textheight]{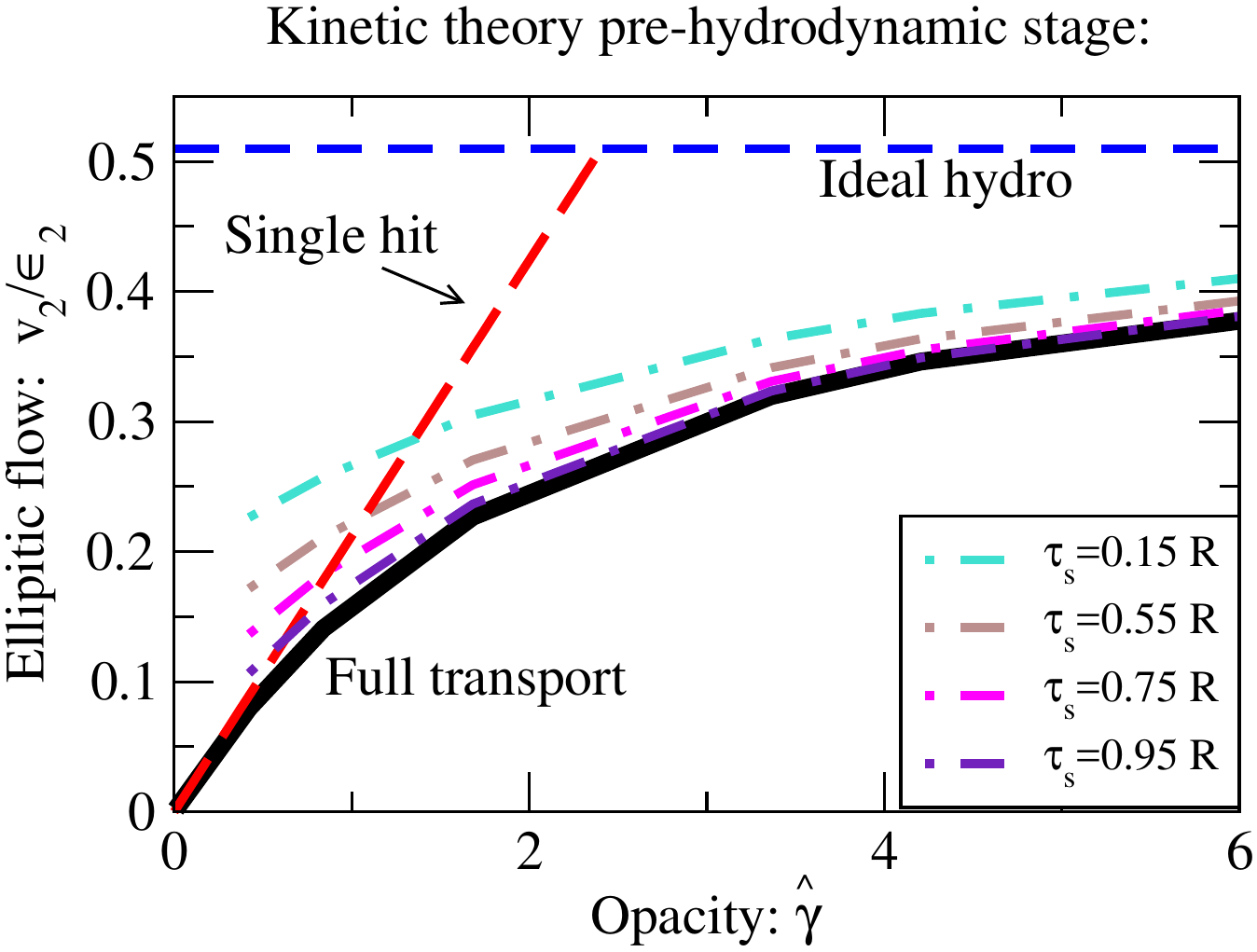}
    \hspace{0.05\textwidth}
    \includegraphics[height=0.2\textheight]{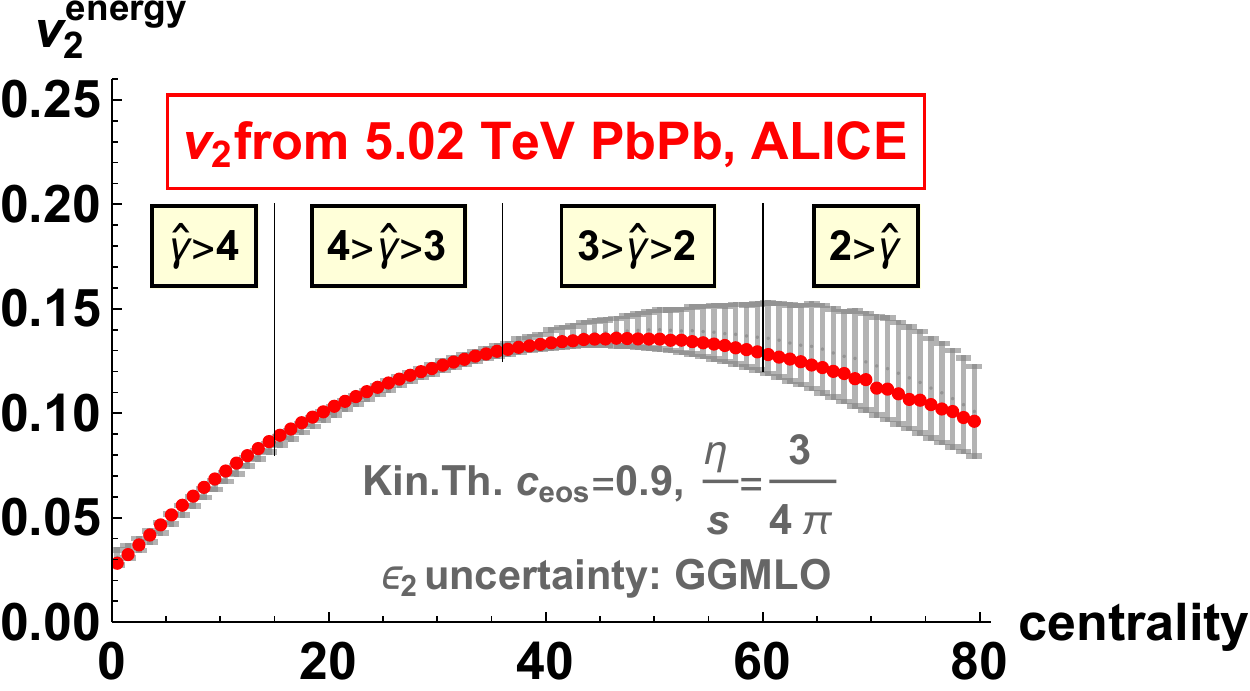}
    \caption{Elliptic flow in the kinetic theory. The left plot shows $v_2$ as a linear response~\cite{Kurkela:2018qeb} and the right plot  (grey band) shows $v_2$ in comparison with data~\cite{Kurkela:2019kip}}
    \label{fig:v2}
\end{figure}

The linear response of $v_n/\epsilon_n$ with $\epsilon_n$ (spatial) eccentricities and $v_n$ (transverse energy) flow coefficients has been evaluated for $n=2$~\cite{Kurkela:2018qeb,Kurkela:2019kip} (the left plot in Fig. \ref{fig:v2}) and $n=3$~\cite{Kurkela:2019kip} using this kinetic theory. Based on these results, a phenomenological study of flow in pA and AA collisions have been conducted (see $v_2$ in AA collisions in the right plot of Fig. \ref{fig:v2}). Unlike central AA collisions, non-hydrodynamic modes are found to dominate in pA collisions by confronting this theory with experimental data. This, hence, calls for a further investigation on the dominance of hydrodynamic modes in hydrodynamic models used for small colliding systems (see~\cite{Nagle:2018nvi} for a recent review on these models). Recently, the nonlinear response of flow coefficients has been calculated and the connection of hydrodynamization in systems with arbitrary transverse profiles to collective flow has been investigated using this kinetic theory~\cite{Kurkela:2020wwb}. 

%\bibliographystyle{JHEP}
%\bibliography{hp2020.bib}

\end{document}